\documentclass[aps,twocolumn,groupedaddress,prl,bibnotes]{revtex4}
\usepackage{graphicx}
\usepackage{revsymb}
\begin{document}


\title{Weak localization in GaMnAs: evidence of impurity band transport}

\author{L.~P. Rokhinson}
\author{Y. Lyanda-Geller}
\affiliation{Department of Physics, Purdue University, West Lafayette, Indiana
47907 USA\\
Birck Nanotechnology Center, Purdue University, West Lafayette, Indiana 47907
USA}

\author{Z. Ge}
\author{S. Shen}
\author{X. Liu}
\author{M. Dobrowolska}
\author{J.~K. Furdyna}
\affiliation{Department of Physics, University of Notre Dame, Notre Dame,
Indiana 46556 USA}

\date{Submitted on February 28, 2007; resubmitted on April 23, 2007}

\begin{abstract}
We report the observation of negative magnetoresistance in the ferromagnetic
semiconductor GaMnAs at low temperatures ($T<3$ K) and low magnetic fields ($0<
B <20$ mT). We attribute this effect to weak localization. Observation of weak
localization provides a strong evidence of impurity band transport in these
materials, since for valence band transport one expects either weak
anti-localization due to strong spin-orbit interactions or total suppression of
interference by intrinsic magnetization. In addition to the weak localization,
we observe Altshuler-Aronov electron-electron interactions effect in this
material.
\end{abstract}


\maketitle

Dilute magnetic semiconductors (DMS) form a bridge between  conventional
ferromagnetic materials and semiconductors, with the promise of electrostatic
tailoring of magnetic properties\cite{jungwirth06rmp}. If enabled to operate at
room temperature, the DMS materials will play a central role in the rapidly
developing field of spintronics, with applications ranging from sensors to
memories and quantum computing. In Mn-based DMSs such as
GaMnAs\cite{ohno96,ohno98,baxter02} the ferromagnetism is carrier--mediated, so
that their magnetic properties are tightly related to the nature of electronic
transport.

The principal unresolved issue in the physics of GaMnAs concerns the roles of
valence and impurity bands. The Zener model of ferromagnetism (which becomes
equivalent to the RKKY approach) has been proposed by Dietl\cite{dietl97,
dietl00}, and developed by others \cite{jungwirth99,jain01} based on the
assumption of hole transport in the valence band in this and related materials.
Alternatively, it has been suggested that the holes in GaMnAs reside in the
impurity band\cite{bercui01,alvarez03,mahadevan04}. Recent optical studies
provide strong evidence of impurity band formation\cite{sapega05,burch06}.
Understanding the origin of electronic states participating in transport -
which bear on the physical origin of ferromagnetism in III-Mn-V alloys -
clearly constitutes the key to achieving higher $T_c$ in DMSs.

In this letter we demonstrate that low-temperature conduction in GaMnAs is
inconsistent with valence band transport. We observe a peak in
magnetoresistance at very small magnetic fields ($B<20$ mT), which is
independent  of orientation of $B$ with respect to the ferromagnetic easy axis
and to the direction of the electric current. The peak appears below 3.4 K and
increases at lower temperatures. We attribute this effect to the anomalous
negative magnetoresistance of the Aharonov-Bohm (AB) origin
\cite{altshuler80a,kawabata80}. The shape and magnitude of the peak is
consistent with weak localization (WL)\cite{abrahams79,gorkov79} in a three
dimensional (3D) conductor with weak spin-orbit interaction. Holes in the
valence band, on the contrary, experience strong spin-orbit interaction, which
would lead to weak anti-localization (positive
magnetoresistance)\cite{altshuler81,lyanda-geller98a} in the absence of
ferromagnetic order or in suppression of interference effects below $T_c$. In
addition to WL we observe a field-independent increase of resistance at $T<8$
K, a signature of Altshuler-Aronov (AA) electron-electron interaction effect on
resistivity\cite{altshuler79}. Such temperature dependent AA contribution is
almost an order of magnitude larger than the magnitude of the magnetoresistance
peak, as it should be in conventional 3D disordered conductors.

The GaMnAs wafers were grown by molecular beam epitaxy (MBE) on semi-insulating
(001) GaAs substrates. Prior to GaMnAs deposition a 120 nm GaAs buffer was
grown at 590$^{\circ}$C, followed by a 2 nm GaAs buffer grown at
275$^{\circ}$C. 100 nm of Ga$_{1-x}$Mn$_{x}$As was deposited next, with
$x=0.02, 0.05, 0.65$, and 0.08. We will refer to those as 2\%, 5\%, 6.5\% and
8\% Mn samples. The Curie temperatures for these wafers are in the range
$60\textrm{\ K}<T_c<100$ K. The measurements were performed on large Hall bars
(a few mm) oriented along the [110] crystallographic direction. Longitudinal
and Hall resistances were measured using the standard four-probe lock-in
technique with 10 nA excitation current in a dilution refrigerator ($0.05-1.2$
K) and in a pumped $^4$He system ($1.2-300$ K). Magnetic fields in the dilution
refrigerator were generated by a home-made two-axis magnet, which in
combination with a rotator allows us to point the magnetic field in an
arbitrary direction. In the following $B_{\bot}$ refers to the field oriented
normal to the sample surface ($B_{\bot}||[001]$); and for in-plane orientations
the field direction will be indicated explicitly (for example
$B_{[110]}||[110]$).

Temperature dependence of resistivity at zero magnetic filed is plotted in
Fig.~\ref{t-dep} for 5\% Mn and 6.5\% Mn samples. As the temperature is
decreased in the paramagnetic phase, the resistance increases and reaches a
maximum around $T_c$, which can  be attributed to the enhanced spin disorder
scattering. Deep in ferromagnetic phase ($T\ll T_c$) the 2\% Mn sample becomes
insulating (resistivity exhibits hopping transport). In contrast, samples with
$>4\%$ Mn show metallic behavior at low temperatures. However, the resistivity
does not saturate as $T\rightarrow0$, but reaches a minimum at $\sim 8$ K and
then slowly increases as the temperature decreases down to the lowest $T=30$
mK. We do not expect any resistance change due to ferromagnetic ordering at
these temperatures, since thermodynamically magnetization reaches 99\% of its
$T=0$ value already at $T=0.2T_c\approx10$ K. Moreover, an increase in
ferromagnetic order should reduce the resistivity, as in the 8 K$<T<T_c$
temperature range.

\begin{figure}[t]
\def\ffile{t-dep}
\includegraphics[scale=0.9]{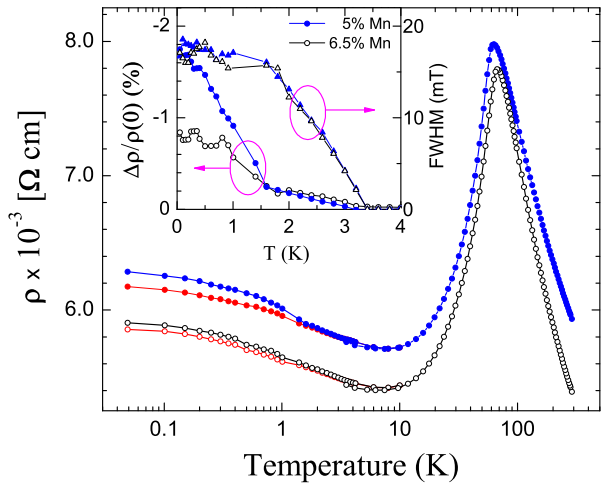}
\caption{Temperature dependence of resistivity at $B=0$ plotted for samples with 5\%
Mn (solid dots) and 6.5\% Mn (open dots). Red point are measured at
$B_{\bot}=30$ mT. In the inset the height
$[\rho(30\mathrm{mT})-\rho(0)]/\rho(0)$ (round symbols) and the full width at
half maximum (FWHM) (triangular symbols) of magnetoresistance peak seen in
Fig.~\ref{magres} are plotted as a function of temperature.}
\label{\ffile}
\end{figure}

\begin{figure}[t]
\def\ffile{magres}
\includegraphics[scale=1.05]{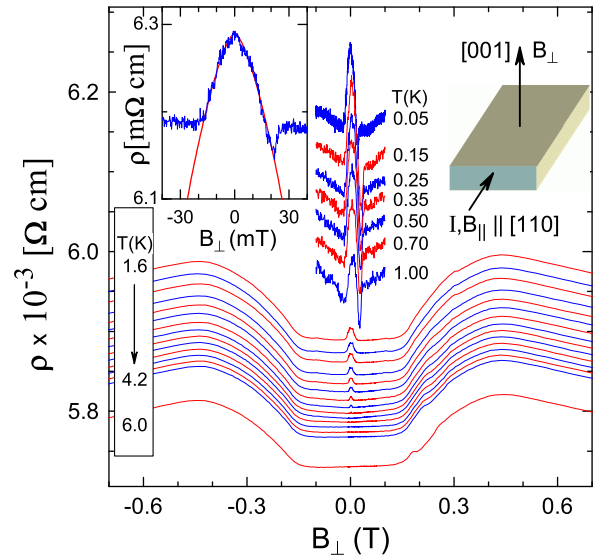}
\caption{Magnetoresistance in 5\% Mn sample plotted at different temperatures $0.05$
$<T<6$ K. Between 1.6 K and 4.2 K $\rho(B)$ is plotted at 0.2 K intervals. In
the inset zero-field peak at 50 mK is enlarged. The red curve is a fit with
$\rho(B)-\rho(0)\propto B^{2}$.}
\label{\ffile}
\end{figure}

\begin{figure}[t]
\def\ffile{magvar1}
\includegraphics[scale=1.2]{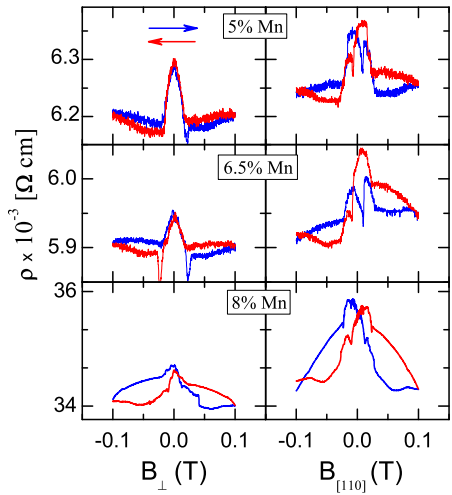}
\caption{Magnetoresistance plotted for 3 samples with different Mn concentrations for
normal ($B_{\bot}$) and in-plane ($B ||[110]$) magnetic fields. Blue (red)
curves are for magnetic field sweeps up (down). Data for
$B_{\bot}$($B_{[110]}$) were measured at 66 mK (23 mK).}
\label{\ffile}
\end{figure}

Weak enhancement of resistivity at low temperatures in disordered conductors is
usually associated with quantum interference and/or interaction effects. Such
interference effects are sensitive to external magnetic fields, and we indeed
observe that the application of a small field of $B_{\bot}=30$ mT reduces
resistivity at $T<3.4$ K. The difference in resistivity
$\rho(30\textrm{mT})-\rho(0)$ is plotted in the inset of Fig.~\ref{t-dep}. This
zero-field enhancement of resistivity reveals itself as a narrow peak in
magnetoresistance. In Fig.~\ref{magres} magnetoresistance at different $T$ is
plotted for an out-of-plane magnetic field $B_{\bot}$. At $T>3.4$ K there is no
change in resistivity in the $-100$ mT$<B_{\bot}<100$ mT range. At $T\approx
3.4$ K a peak appears at $B_{\bot}=0$. The height of the peak then gradually
increases as the temperature decreases, and approaches 1-2\% of the overall
resistivity at $T=30$ mK.  The peak width also increases with decreasing $T$,
and almost saturates for $T<1$~K.

The height and shape of the zero-field peak is independent of the orientation
of magnetic field. In Fig.~\ref{magvar1} we plot magnetoresistance as a
function of the out-of-plane $B_{\bot}$ and in-plane $B_{[110]}$ fields for
three samples with different Mn concentrations. The overall shape of
magnetoresistance is very different for the two field orientations, and
exhibits a hysteretic behavior. The
zero-field peak, however, has no hysteresis and has a similar height and width
for both field orientations, which suggests that its origin is not related to
ferromagnetic ordering. This feature is emphasized in Fig.~\ref{phe}, where
both $R_{xx}$ and the planar Hall effect (PHE) are plotted for different
orientations of the in-plane magnetic field. The jumps in PHE indicate
switching of magnetic domains, which produce corresponding spikes in
magnetoresistance, but do not change the overall shape of the zero-field peak.
We also observe a zero-field enhancement of the PHE, similar to the peak in
magnetoresistance. This behavior is consistent with the PHE resulting from
inhomogeneities of current flow, and reflects the corresponding enhancement of
$R_{xx}$.

We now discuss the experimental data. The only known physics that can explain a low magnetic field magnetoresistance that is independent of the magnetic field
orientation relative to the current and to crystallographic axes is the
phenomenon of weak localization (WL), which leads to anomalous
magnetoresistance arising from the Aharonov-Bohm effect. There are several
distinct experimental features which indicate that the observed effect is
indeed related to WL: (i) The zero-field peak gradually disappears with
increasing temperature as the phase breaking processes intensify, thus
destroying WL. (ii) Similar temperature dependence characterizes also the
magnetic-field-independent background. This behavior is characteristic to the Altshuler-Aronov (AA) electron-electron interactions effect on resistivity that accompanies WL in disordered conductors at low temperatures. Indeed, the AA
effect is not destroyed (to a leading order) by the Aharonov-Bohm magnetic flux passing through electron trajectories, and its magnetic field dependence due to spin arises only in rather strong magnetic fields \cite{lee82a}. Furthermore,
in 3D conductors the AA contribution to magnetoresistance should exceed the
WL contribution by an order of magnitude. The experimentally measured ratios are 4 and 11 for the wafers with 5\% and 6\% Mn respectively. (iii) The value of magnetoresistance is also consistent with the WL physics. Furthermore, from the
suppression of the WL peak at 20 mT we estimate the phase breaking length to be $l_{\phi}\approx 0.1$ microns. This estimate is consistent with the inelastic
phase breaking length extracted from universal conduction fluctuations in
similar materials\cite{wagner06,vila06}. (iv) the shape of the zero-field peak
is consistent with theoretically expected $B^{2}$ dependence (see inset in
Fig.~\ref{magres}).

\begin{figure}[t]
\def\ffile{phe}
\vspace{-0.25in}
\includegraphics[scale=1]{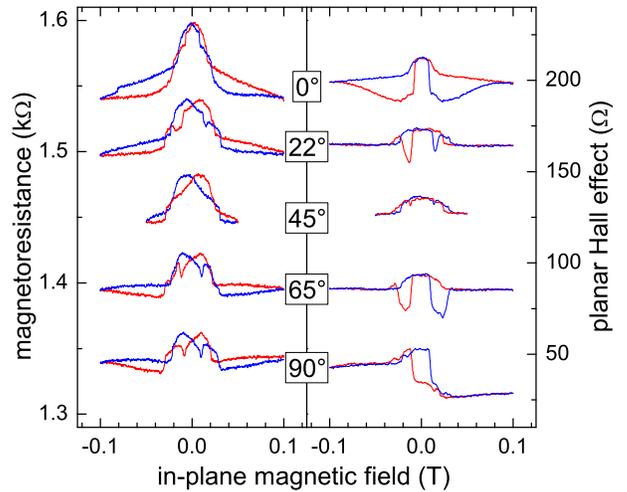}
\caption{Magnetoresistance (left panel) and planar Hall effect (right panel) shown for
different orientations of in-plane magnetic field in the 6.5\% Mn sample. All curves
except $\alpha=90^{\circ}$ are offset for clarity. Angle $\alpha=0^{\circ}$ corresponds
to $B||[110]$, $\alpha=45^{\circ}$ to $B||[100]$ (the easy axis of magnetization), and
$\alpha=90^{\circ}$ to $B||[1\bar{1}0]$. Current $I || [110]$. Blue (red) curves are for
magnetic field sweeps up (down).}
\label{\ffile}
\end{figure}

We thus attribute the zero-field peak in magnetoresistance to the WL effect.
This observation is intriguing, because GaMnAs is a magnetic alloy, so that
magnetic interactions must coexist with WL, which limits their strength.
Furthermore, the negative sign of the observed magnetoresistance brings certain
restrictions on the properties of charge carriers contributing to the
resistivity.

The WL correction to conductivity for charge carriers with spin (angular
momentum) 3/2 can be written as
\begin{equation}
\Delta\sigma\propto-\frac{1}{4}\big(
-S_0+\sum_{i=-1}^1T_{1,i}-\sum_{i=-2}^{2}Q_{2,i}+\sum_{i=-3}^{3}S_{3,i}\big)
\label{MR}
\end{equation}
where $S_0$ is the singlet contribution to conductivity of interfering electron
waves with total spin zero, and $T_{1,i}$, $Q_{2,i}$, $S_{3,i}$ are triplet,
quintuplet and septuplet contributions. They arise from the total angular
momenta 1, 2 and 3 respectively, $i$ being the projection of angular momentum
on the quantization axis.

When orbital (Aharonov-Bohm) effects suppress the interference contributions,
one observes either a negative or a positive magnetoresistance, depending on
the relative importance of the multiplets and singlet. This relative importance
is determined by the strength of spin-dependent interactions. A negative
magnetoresistance requires that spin and spin-orbit scattering are negligible,
that almost no intrinsic spin-orbit interactions present, and that charge
carriers are not affected by a strong Zeeman effect and/or ferromagnetism. If
these conditions are satisfied, the Aharonov-Bohm flux suppresses localization
of electrons, leading to negative magnetoresistance, which  is defined by the
sum of singlet, triplet, quintuplet and septuplet contributions. We note that
when all spin-dependent interactions are absent, each of the multiplets
contributing to WL is equal to the singlet, leading to a localizing correction,
$\Delta\sigma\propto -S_0$, and to negative magnetoresustance. However, if
strong spin-orbit effects completely suppress all multiplet contributions, the
remaining singlet contribution will lead to anti-localization,
$\Delta\sigma\propto \frac{1}{4}S_0$, and, correspondingly, to a positive
magnetoresistance.

GaMnAs is characterized by several magnetic interactions: spins are affected by
average magnetization in the ferromagnetic phase, by scattering off magnetic
fluctuations due to Mn, by domain walls and other magnetization
inhomogeneities, and finally by intrinsic spin-orbit interactions and
spin-orbit scattering. Observation of WL allows us to make conclusions about
dominant hole scattering mechanisms. In particular, we conclude that scattering
by magnetic fluctuations cannot be dominant. Otherwise the singlet and
multiplet terms, both affected by such scattering, would be entirely
suppressed, leading to the absence of interference effects. Thus it is the very
strong positional disorder rather than magnetic scattering that dominates the
scattering mechanism, limiting the mean free path.

In contrast to scattering off magnetic fluctuations, average ferromagnetic
magnetization of GaMnAs must have a profound effect on the interference terms,
entirely suppressing contributions with antiparallel spins in singlet and
multiplet states. If $S_0$ is entirely suppressed, the only contributions to
weak localization arise from multiplets, resulting in negative
magnetoresistance. The remaining magnetic interactions: domain walls and other
smooth magnetic inhomogeneities, and various types of spin-orbit interactions,
can only affect multiplet terms with non-zero projections of angular momentum
\cite{lyanda-geller98}. If these magnetic interactions are weak, then a
negative magnetoresistance can indeed be observed, as seen experimentally.

We can now set a restriction on the origin of carriers that contribute to the
conductivity. If the contributions were coming from valence band holes in GaMnAs, then
strong spin-orbit interactions of total angular momentum of holes with their kinetic
momentum $k$ would result in spin dephasing (scattering) times of the order of the
transport scattering time. This would lead to the total suppression of multiplet terms,
resulting in a positive magnetoresistance similar to that observed in non-magnetic p-type
materials\cite{papadakis02}. In impurity band, there is no well defined $k$ vector and
the aforementioned spin dephasing mechanism is absent. Recent consideration of spin
dephasing for electrons in a shallow impurity band show that dephasing mechanism
effective in the impurity band gives 2-3 orders of magnitude increase of spin dephasing
time compared to conduction band\cite{tamborenea07}. Similar effect on spin dephasing due
to the absence of $k$ is expected for holes in an impurity band compared to valence band.
The suppression is expected to be even stronger in GaMnAs due to the deep nature of Mn
acceptors (112 meV). Thus the spin-orbit effects in the impurity band have only limited
impact on multiplet terms, in agreement with the observed negative magnetoresistance.

Finally, we would like to point out several unusual features observed in our
experiments. Typically a negative magnetoresistance in 3D disordered
nonmagnetic conductors has $B^2$ field dependence at low $B$, which smoothly
evolves into $\sqrt{B}$ at higher $B$ (at $l_m\sim l_{\phi}$, where $l_m$ is
the magnetic length). In our samples, however, instead of such gradual change
of magnetoresistance with field we observe an abrupt suppression of the effect.
A related feature is the $T$-dependence of the width of the magnetoresistance
peak. From $l_m\sim l_{\phi}$ crossover one expects that
the peak will broaden with increasing temperature (since $l_m\propto B^{-1/2}$
and $l_{\phi}\propto T^{-1}$). In our data, however, we observe just the
opposite: the magnetoresistance peak narrows as the temperature increases. We
analyzed several mechanisms which can potentially lead to the suppression of WL
and are enhanced at higher temperatures. In weak magnetic fields the average
spin inside the domains begin to tilt away from the easy axis, and the
resulting spin texture will then act as an effective Berry's phase. This
suppression mechanism \cite{lyanda-geller98} should not be present for the
field aligned along the [100] (i.e., the easy axis) direction. Experimentally,
however, the peak for $B || [100]$ is the same as for the other field
directions, see Fig.~\ref{phe}. Also, the observation of domain switching
within the WL peak rules out the possibility that the suppression of WL is
caused by domain walls.

In conclusion, we have observed an unexpected negative magnetoresistance at small
magnetic fields in GaMnAs, which we attribute to weak localization. We also observe weak
temperature dependence of resistivity which we ascribe to the Altshuler-Aronov
electron-electron interactions effect. The sign of magnetoresistance indicates that
transport in GaMnAs cannot originate from valence band holes, but must be attributed to
holes in the impurity band. Observation of interference effects in resistivity at high
($>4$\%) Mn concentrations indicates that the hole transport is diffusive.

The work was supported by NSF under the grants ECS-0348289 (LR) and DMR-0603752
(Notre Dame).

\bibliography{rohi}

\end{document}